\documentclass[12pt,preprint]{article}
\usepackage{epsf}
\newcommand{\eqb}{\begin{equation}}
\newcommand{\eqe}{\end{equation}}
\newcommand{\dmb}{\begin{displaymath}}
\newcommand{\dme}{\end{displaymath}}
\newcommand{\pd}{\partial}

\newcommand{\eab}{\begin{eqnarray}}
\newcommand{\eae}{\end{eqnarray}}
\newcommand{\ra}{\right\rangle}
\newcommand{\la}{\left\langle}

\newcommand{\be}{\begin{equation}}
\newcommand{\ee}{\end{equation}}

\newcommand{\La}{\Lambda}
\def\lsim{\mathrel{\raise.3ex\hbox{$<$\kern-.75em\lower1ex\hbox{$\sim$}}}}
\def\gsim{\mathrel{\raise.3ex\hbox{$>$\kern-.75em\lower1ex\hbox{$\sim$}}}}
\def\Li2{{\rm Li}_2}

\begin{document}
\begin{titlepage}
\begin{flushright}
MPI-PhT 2002-41 \\
\end{flushright}
\vspace{0.6cm}

\begin{center}
\Large{{\bf The curvaton as a Bose-Einstein condensate 
of chiral pseudo Nambu-Goldstone bosons}}

\vspace{1cm}

Ralf Hofmann

\end{center}
\vspace{0.3cm}

\begin{center}
{\em Max-Planck-Institut f\"ur Physik\\ 
Werner-Heisenberg-Institut\\ 
F\"ohringer Ring 6, 80805 M\"unchen\\ 
Germany}
\end{center}
\vspace{0.5cm}

\begin{abstract}

Explaining cosmic inflation by the effective dynamics of an $SU(N_C)$ pure gauge 
theory of scale $\La\sim 10^{-6}\,M_p$ is lacking an explanation of the 
(Gaussian) spatial curvature perturbations needed to seed the 
formation of large-scale structure 
after inflation. In this work it is demonstrated how 
fundamentally charged fermions of mass $m_F\sim 10^{-12}\,\La$, 
whose approximate chiral symmetry is spontaneously broken 
during inflation, can cure this shortcoming. The associated gas of 
weakly interacting pseudo Nambu-Goldstone bosons (PNGB) undergoes 
Bose-Einstein (BE) condensation well before the 
end of inflation. This causes the occurence of 
condensed light scalar fields effectively acting as a curvaton. Fermions may 
also be charged under a gauge group $G$ with a weak coupling $g$. 
Since fermions charged under $SU(N_C)$ are confined after inflation 
the decay of Nambu-Goldstone bosons, 
which reside in the condensates and in the PNGB radiation 
generated at reheating, is mainly into fermions {\sl solely} charged 
under $G$. The associated decay rate $\Gamma$ is estimated using 
PCAC and large $N_C$ counting. PNGB decay takes place during radiation domination 
after cosmological scales have entered the horizon. 
Neglecting the effects of the spontaneous breaking of G induced by 
the BE condensation, it is demonstrated that the observed 
spectrum of spatial curvature 
perturbations $P^{1/2}_\xi\sim 5\times 10^{-5}$ is compatible with 
$g\sim 10^{-3}$, a ratio $r_{\tiny{dec}}\sim 10^{-4}$ of 
curvaton to radiation energy at PNGB decay, 
and a curvaton effective equation 
of state $p_{\tiny{curv}}=-0.9\,\rho_{\tiny{curv}}$.

\end{abstract} 

\end{titlepage}

\section{Introduction}

Cosmic inflation is an attractive approach to 
the horizon, flatness, monopole, and 
homogeneity problem \cite{Inflation}. It is of 
primary interest to find an answer to the question how 
large-scale structure is seeded and how the 
anisotropies of the cosmic microwave background at 
large angular resolution 
come about, see \cite{Lythbook} 
and references therein. More recently, 
string inspired alternatives to inflation have 
been proposed where cosmology is driven by the dilaton field \cite{Veneziano} or 
by a radion field \cite{Ovrut} which determines the 
distance between two parallel and four dimensional 
boundary branes of a five dimensional bulk. 
In their minimal versions both models, however, have been claimed to not be 
able to generate a scale-invariant spectrum of 
adiabatic fluctuations \cite{Lythpbb-ep,Brandpbb-ep}.

The explanation of inflation as well as the description of 
adiabatically generated primordial curvature perturbations by the real-time dynamics 
of one or more slowly rolling, 
real, and gravitationally minimally coupled 
scalar fields is an effective approach. Inflationary models of this type are hardly 
constrained by observation, and hence a large 
amount of arbitraryness is inherent to the model building. 
To attenuate this unsatisfactory fact and to build on a solid principle it was proposed 
in Ref.\,\cite{HofKeil} that the strong dynamics driving very early comology is due to 
an $SU(N_C)$ gauge symmetry with large $N_C$. 

Effectively, the thermal behavior of the pure 
gauge theory is modelled by quasi-particle excitations 
over a nontrivial classical background. An adjoint Higgs model at very high temperatures is 
smoothly connected to a dynamical Abelian projection at high temperatures which, in turn,  
reduces to a $Z_{N_C}$ symmetric model for $N_C-1$ complex scalar fields $\phi_k$ 
at $T\lsim\frac{\Lambda}{2\pi}$ \cite{Hofmann03}. The latter model has a 
vanishing vacuum energy and therefore terminates inflation. 
Here $\La$ denotes the dynamical scale of 
the $SU(N_C)$ theory which also sets the scale for inflation. 
In this paper we focus on the Abelian 
and the center phase. A finite modulus $|\phi_k|<\La$ embodies the $T$ 
dependent condensation of magnetic monopoles 
whereas $|\phi_k|\sim\La$ describes the condensation of center vortices. 
Qualitatively, this picture is supported by lattice 
simulations of an $SU(2)$ gauge theory at finite temperature \cite{deForcrand}. 
The potential $V$ of our effective description is constructed to admit topologically 
nontrivial, periodic  
Bogomoln'yi-Prasad-Sommerfield (BPS) \cite{BPS} saturated 
solutions to the Euclidean $\phi$-field equations along 
a compact time dimension of length $\beta\equiv 1/T$ \cite{HofKeil,Hof}.     

The effective gauge theory description implies radiation 
domination for $M_P>T\gg\frac{\Lambda}{2\pi}$ which 
runs into vacuum domination for $T\sim\frac{\Lambda}{2\pi}$. The latter regime is 
terminated by a transition to the phase of condensed center vortices. 
In the absence of fermions no (pseudo) 
Nambu-Goldstone moduli are produced 
in this transition since $Z_{N_C}$ is a 
discrete symmetry. Heavy vector bosons, which survive inflation, are
stable and possibly contribute to today's 
cold dark matter. In the 
presence of fundamentally charged, low-mass 
fermions a kinematically possible, postinflationary decay of vector
bosons into $Z_{N_C}$-charged {\sl chiral} 
pseudo Nambu-Goldstone bosons (PNGB) 
should be strongly suppressed by higher dimensional operators.  

Since the monopole condensate fluctuates with 
a mass $m_{\phi}\sim \Lambda\gg H_i$ during inflation, 
$H_i$ denoting the (quasi-constant) Hubble parameter, 
the usual mechanism of an adiabatic 
generation of spatial curvature perturbations fails 
in the case of pure gauge dynamics \cite{HofKeil}. 
To propose an alternative mechanism for the generation of spatial
curvature perturbations in the framework of an effective 
gauge theory description of inflation 
is the purpose of the present paper. 
The scenario relies on the curvaton 
mechanism which was discovered 
in Ref.\,\cite{curvaton} and 
re-addressed recently in Ref.\,\cite{LythWands}. 
In its minimal version this mechanism operates with a gravitationally 
minimally coupled, light scalar 
field -- the curvaton -- 
which is decoupled from the dynamics driving inflation. 
During inflation the quantum fluctuation of the 
curvaton convert into isocurvature 
perturbations on super-horizon scales. 
Upon horizon-entry of cosmological scales during the 
radiation dominated epoch after inflation isocurvature perturbations 
get converted into spatial curvature perturbations 
due to an increasing ratio of 
curvaton to radiation energy density. Following the epoch of 
curvaton decay into radiation the 
spatial curvature perturbation remains 
constant in time and is responsible for the formation 
of large-scale structure \cite{Lythbook}. In our gauge theory framework 
it is suggestive that the existence of a 
curvaton has to do with the dynamical 
breakdown of a continuous, global symmetry 
due to strong gauge interactions close 
to the transition to the confining phase. 
This symmetry is present if we allow 
for fundamentally charged, chiral fermions 
to enter the stage. In this case, 
the spontaneous breakdown of 
chiral symmetry {\sl during} 
inflation releases a gas of chiral PNGB which 
undergoes Bose-Einstein (BE) condensation after a 
sufficient number of e-foldings. The resulting 
condensate acts as a curvaton and decays into fermions solely charged 
under a weakly coupled gauge group $G$. 
 
The outline of the paper is as follows: In Section 2 we introduce our model for 
the relevant fields during inflation, discuss BE condensation 
of chiral PNGB in an expanding universe, and use a 
simplified, effective description of the condensate to estimate the mass 
of fundamental fermions from the requirement of 
scale invariance of the Gaussian isocurvature 
perturbations generated during inflation. The postinflationary 
generation of curvature perturbations and the 
subsequent decay of PNGB are discussed in Section 3, 
and the consistency of various assumptions is checked. 
There is a short summary in Section 4.

\section{Chiral symmetry breaking and 
BE condensation during inflation}

\subsection{Set-up and general considerations}

In this paper we consider the dynamics of 
matter that is not directly related to the Standard Model 
of particle physics and 
its supersymmetric extensions.  

We consider $N_F$ flavors of asymptotically free fermions 
of degenerate mass $m_F$ which transform 
fundamentally under an $SU(N_C)$ gauge theory of scale $\La$. 
At one loop accuracy we have $N_C>\frac{2}{11}\,N_F$. Assuming 
$m_F\ll\La$, there is an approximate chiral 
symmetry $U(N_F)$. For this symmetry group to possess 
nontrivial, nonanomalous subgroups we impose $N_F>1$. 
To have a sensible $1/N_C$ 
counting we assume $N_F\ll N_C$. The fermions are taken to be fundamentally charged under 
an additional, much smaller gauge group $G$ inducing weakly coupled dynamics with a 
gauge coupling constant $g\lsim 1$\footnote{The consistency of this
assumption will be shown in Section 4.2.}. 
The total gauge group therefore is $SU(N_C)\times G$. 
We will refer to the fermions characterized above as $F$-fermions. 
Assuming that the vacuum energy of the theory is dominated by that of the 
pure glue sector, one obtains an effective groundstate description as in 
\cite{HofKeil,Hof}. 
The magnetic coupling constants of the $SU(N_C)$ gauge bosons 
to the monopole condensate, which is 
described by a single, complex mean field $\phi$, are 
determined by the root structure of $SU(N_C)$ \cite{Suzuki}. 
For simplicity we assume all magnetic couplings and 
hence all gauge boson masses to be equal.

In addition, $K$ flavors of fermions with a 
mass comparable to $m_F$ are assumed. These fermions be fundamentally charged {\sl solely} 
under $G$, and in the following we will refer to them 
as $K$-fermions. Einstein gravity is assumed to 
be minimally coupled to this matter. 

For large $N_C$ the critical temperature for the transition 
to the confining $Z_{N_C}$ symmetric phase is 
$T\sim\frac{\Lambda}{2\pi}$ \cite{HofKeil}. 
If we assume that charge 
parity C is a good symmetry in the deconfining phase then close to the transition 
the ratio $R$ of radiation pressure to vacuum pressure is given as 
\eqb
\label{est}
R=(2\pi)^{-4}\times 12\left[N_F+\frac{1}{N_C}\left(K\,\mbox{dim}[\mbox{fund}(G)]+
\frac{1}{2}\mbox{dim}[\mbox{adj}(G)]\right)\right]\,.
\eqe
Eq.\,(\ref{est}) derives from the facts that the gauge bosons of the dynamical Abelian projection 
decouple close to the transition to the center phase \cite{Hofmann03} and that there are 
$4(N_F N_C+K\mbox{dim}[\mbox{fund}(G)])$ chiral fermions and $2\,\mbox{dim}[\mbox{adj}(G)]$ massless 
gauge bosons of the gauge group $G$.  

Hence the ground state dominates the cosmic evolution 
(quasi-exponential expansion, quasi de Sitter geometry) 
close to the phase transition if $K$ and $N_F$ 
are not too large. 
For example, $N_C=34$, $G=SU(2)$, 
and $K=N_F=3$ yields $R\sim3.6$\,\%.    

Quenched QCD lattice simulations seem to indicate that the critical temperatures 
for chiral restoration is slightly higher than the 
deconfinement transition \cite{lattice}. Including dynamical fermions 
and appealing to the method of imaginary chemical potentials 
it was found recently in Ref.\,\cite{Fodor} that 
$N_C=2,3$ QCD has no confinement phase transition at all 
but rather a level crossing. This is 
consistent with \cite{Hof} where the nonexistence of a 
tachyonic regime in the model effective 
potential was observed for $N_C<8$. However, the classical arguments of Ref.\,\cite{Hof} 
are only applicable in the limit of large $N_C$, and 
therefore this consistency at small $N_C$ may be pure coincidence. 
We nevertheless take the results of \cite{lattice} as a motivation to 
assume that in the framework of Ref.\,\cite{HofKeil}, where $N_C=34$ was found from the condition that
inflation is terminated after 60 e-foldings, the 
chiral transition occurs {\sl during} the de Sitter 
stage. This transition naturally 
generates a radiation of PNGB.

Our intuition about inflationary cosmology seems to forbid 
the condensation of PNGB since the 
exponential expansion of the universe tears 
particles apart and allows for causal contact only 
within a quasi-constant horizon. This, however, may not be a valid concern: 
The total particle number as 
measured at the instant of emission is only conserved on super-horizon scales. 
A particle being emitted at some instant $t_0$ is in a 
momentum eigenstate normalized to 
the Hubble volume. At some later time $t_1>t_0$ this volume 
has expanded by a factor 
$\exp[3H(t_1-t_0)]$ decreasing the contribution of the 
particle to the total 
density by a factor $\exp[-3H(t_1-t_0)]$. Since the position of 
a particle is irrelevant for the formation of condensates we may still think in terms 
of conserved particle numbers. 
Therefore, the relevant patch of the inflating universe, which coincides with the observable universe
at emission, may be associated with 
a box of exponentially 
increasing volume containing a gas of 
exponentially decreasing temperature. Bose-Einstein (BE) 
condensation is a pure quantum effect which needs no 
interaction betweeen the participating 
particles \cite{deGroot}. Hence a comparison of expansion 
and interaction rate again is irrelevant for the formation of condensates. Therefore, 
BE condensation may be expected to take place after a 
sufficiently long time during the de Sitter epoch.    

The time scale for the duration of the chiral transition is set by $f_{\tiny{\mbox{PNGB}}}\sim\La$. 
Since $\La\sim\frac{M_P}{\La}H_i\gg H_i$ this transition takes place on a much smaller 
time scale than the inflationary 
expansion of the universe, and hence the chiral transition may be viewed as 
instantaneous. We assume that the spectrum of PNGB generated at the instant of emission 
is of the Bose-Einstein type with temperature $T\sim\frac{\La}{2\pi}$ and 
chemical potentials $\mu_e^\alpha$. In a given gauge 
the $\mu_e^\alpha$ correspond to charges $Q_e^\alpha$ within the horizon arising from the 
inhomogeneities in the charge 
densities $n^\alpha$ which are defined as the zero components of 
$G$ gauge covariantly conserved currents $j^\alpha_\mu$. 
Because the coupling $g$ is small and 
since the chiral interaction strength between PNGB is parametrically 
counted in powers of $\frac{T}{4\pi\,f_{\tiny{\mbox{PNGB}}}}$ 
($f_{\tiny{\mbox{PNGB}}}\sim\La$) and at the instant of 
PNGB emission $T\sim\frac{\La}{2\pi}$ the treatment of the PNGB as a Bose gas of 
noninteracting particles is justified.

\subsection{BE condensation in an expanding universe}

We have to distinguish two cases. 
A chiral PNGB particle is either 
(a) charged or (b) neutral under the group $G$. 
Condensation of charged PNGB can only 
occur if there are, in a given gauge, nonvanishing 
charge densities $n^a$ within the de Sitter horizon
being produced during the chiral transition. Here $\alpha$ runs through the representation 
of $G$ which acts on the charged PNGB being made of a pair of an 
F-fermion and F-antifermion. There are $N_F^2$ 
species of PNGB for each $G$ charge. We do not 
carry this factor along in the following discussion. 
Moreover, we will omit the label $\alpha$ in the following 
since for generic gauge groups many of 
the $n^\alpha$ can be gauge-rotated away. 
 
Let us first consider case (a). A $G$ 
charge $Q_e$ within the inflationary 
horizon at the instant of the chiral 
transition is associated with a chemical 
potential $\mu_e$. We have 
\eab
\label{charge}
n_e&=&\frac{1}{2\pi^2}\int_{0}^\infty dp\,
p^2\times\nonumber\\ 
& &\left[\frac{1}{\exp[\frac{2\pi}{\La}(\sqrt{p^2+m_{\tiny{\mbox{PNGB}}}^2}-\mu_e)]-1}-
\frac{1}{\exp[\frac{2\pi}{\La}(\sqrt{p^2+m_{\tiny{\mbox{PNGB}}}^2}+\mu_e)]-1}\right]\,.
\eae
Let $x$ denote the number of e-foldings the scale factor $a$ undergoes 
from the instant of the chiral transition to the 
first occurence of the charged BEC. Then the charge density  
$n_e$ relates to the charge density $n_0$ at the onset of condensation 
as
\eqb
\label{chrcon}
n_0=\exp[-3x]\,n_e\,.
\eqe
In an ideal, nonrelativistic, and bosonic quantum gas of identical particles 
the charge density $n_0$, in turn, determines the critical 
temperature $T_0$ for the onset of BE condensation as
\cite{deGroot}
\eqb
\label{n0j}
n_0=2.612\times\left(\frac{T_0 m_{\tiny{\mbox{PNGB}}}}{2\pi}\right)^{3/2}\,.
\eqe
Combining Eqs.\,(\ref{charge}),(\ref{chrcon}), and (\ref{n0j}), we derive
\eab
\label{T01}
T_0&=&\frac{2\pi}{m_{\tiny{\mbox{PNGB}}}}\,\exp[-2x]\,\left(\frac{1}{2.612\times 2\pi^2}
\int_{0}^\infty dp\,
p^2\times\right.\nonumber\\ 
& &\left.\left[\frac{1}{\exp[\frac{2\pi}{\La}(\sqrt{p^2+m_{\tiny{\mbox{PNGB}}}^2}-\mu_e)]-1}-
\frac{1}{\exp[\frac{2\pi}{\La}(\sqrt{p^2+m_{\tiny{\mbox{PNGB}}}^2}+\mu_e)]-1}\right]\right)^{2/3}\,.\nonumber\\
&& 
\eae
Eq.\,(\ref{T01}) relates the critical temperature $T_0$ to the 
chemical potential at emission $\mu_e$ and the 
number of e-foldings $x$ needed for the onset of condensation.

After condensation there must be sufficient inflation to generate classical, Gaussian 
isocurvature fluctuations of the BEC's which are converted to curvature 
perturbations after inflation (see Section 3.1). Therefore, 
BE condensation must be completed within a number of e-foldings 
considerably smaller than the total of 60 or so 
inflationary e-foldings required to solve the horizon problem. 

It is the finiteness of the scale $m_{\tiny{\mbox{PNGB}}}$ that makes it likely to cause a finite 
charge density $n_e$ during the chiral transition. Thus we assume $\mu_e$ 
to be comparable to $m_{\tiny{\mbox{PNGB}}}$. Moreover, we assume that 
$m_{\tiny{\mbox{PNGB}}}$ also sets the scale 
for $T_0$. There is no strict reason why this should be so. 
However, $m_{\tiny{\mbox{PNGB}}}\equiv \mu_e\equiv T_0$ leads to 
an acceptable number of e-foldings for 
the completion of BE condensation if we 
use $\La\sim 10^{13}$\,GeV as in Ref.\,\cite{HofKeil}.
 
Let us make this more explicit. As a result of Section 2.4 
we take $m_{\tiny{\mbox{PNGB}}}=10^{-6}\,\La$. Inserting $m_{\tiny{\mbox{PNGB}}}\equiv \mu_e\equiv T_0$, 
$\La=10^{-6}\,M_p$, $m_{\tiny{\mbox{PNGB}}}=10^{-6}\,\La$ into Eq.\,(\ref{T01}) 
and varying $\mu_e$ between $\mu_e=0.1 m_{\tiny{\mbox{PNGB}}}$ and $\mu_e=0.9 m_{\tiny{\mbox{PNGB}}}$, 
the number of e-foldings $x$ needed for the onset of BE condensation 
varies between $x=7.5$ and $x=8.2$. Hence, there is a 
good insensitivity of $x$ with respect to variations in 
$\mu_e$. Note that with $H_i\sim\frac{\La^2}{M_p}\sim 10^{-12} M_p$ 
we satisfy the condition $H_i<10^{-5}\,M_p$ which 
is inferred from the cosmic microwave 
background limit on gravitational 
waves. Here $M_p$ denotes the Planck 
mass $M_p\sim 10^{19}$\,GeV in contrast 
to a reduced Planck mass $M_P$ introduced below. 

To estimate the number of e-foldings $x_c$ needed to 
complete condensation we consider the following ratio as it was 
derived in Ref.\,\cite{deGroot}
\eqb
\label{ratioc}
\frac{n_c}{n_{tot}}=1-\left(\frac{T}{T_0}\right)^{3/2}\,.
\eqe
Here, $n_{tot},n_c$ denote the total density of PNGB and the density 
fraction of particles residing in the condensate, respectively. We decompose 
$n_{tot}(x)=n_c(x)+\bar{n}(x)$, where $\bar{n}(x)$ refers to the fractional 
density of particles in the gas phase after $x$ 
e-foldings counted from the onset of condensation. 
For $T\le m_{\tiny{\mbox{PNGB}}}$ temperature scales nonrelativistically, that is 
$T=T_0\,\exp[-2x]$. Substituting this 
into Eq.\,(\ref{ratioc}), we obtain
\eqb
\label{barn}
\bar{n}(x)=n_c(x)\left(\exp[3x]-1\right)^{-1}\,.
\eqe
Since particles in the 
condensate have no spatial momentum 
their density $n_c$ does not scale with $x$ 
as a {\sl direct} consequence of the 
expansion of the universe. 
Because of Eq.\,(\ref{barn}), which describes the $T$ dependence of 
the condensate fraction, there is, however, 
an {\sl indirect} scaling. This can be seen as follows. 

Changing $x\to x+dx$, the variation $d\bar{n}$ is due to gas particles going into the 
condensate on the one hand and due to a decrease of gas density 
because of de Sitter expansion on the other hand. 
Hence, we have 
\eqb
\label{dbarn}
d\bar{n}=-(n^\prime_c+3\bar{n})dx\,,
\eqe
where $n^\prime_c\equiv \frac{dn_c}{dx}$. Replacing $x\to x+dx$ in 
Eq.\,(\ref{barn}), using Eq.\,(\ref{dbarn}), and comparing 
coefficients at $dx$, we obtain
\eqb
\label{ncp}
n^\prime_c=3\frac{\exp[-6x]}{1-\exp[-3x]}\,n_c\,.
\eqe
The solution to Eq.\,(\ref{ncp}) reads
\eqb
\label{solncp}
n_c(x)=n_{c,\infty}\,\exp\left[\exp[-3x]\right]\,(1-\exp[-3x])\,.
\eqe
Eqs.\,(\ref{barn}),(\ref{dbarn}) can also be solved for $n_c,dn_c$, respectively. 
Repeating the above steps in this case yields
\eqb
\label{nbarnp}
\bar{n}^\prime=-3\frac{\exp[-3x]}{1+\exp[3x]}\,\bar{n}\,.
\eqe
The solution to Eq.\,(\ref{nbarnp}), subject to the initial condition 
$\bar{n}(0)=n_0$, is
\eqb
\label{nbarnsol}
\bar{n}(x)=\frac{n_0}{e}\,\exp\left[\exp[-3x]-3x\right]\,.
\eqe
From Eq.\,(\ref{barn}) we finally obtain
\eqb
\label{nc0}
n_{c,\infty}=\frac{n_0}{e}\,.
\eqe
Fig.\,1 shows a plot of $n_c(x)/n_{c,\infty}$. 
\begin{figure}
\begin{center}
\leavevmode
\leavevmode
\vspace{7cm}
\includegraphics{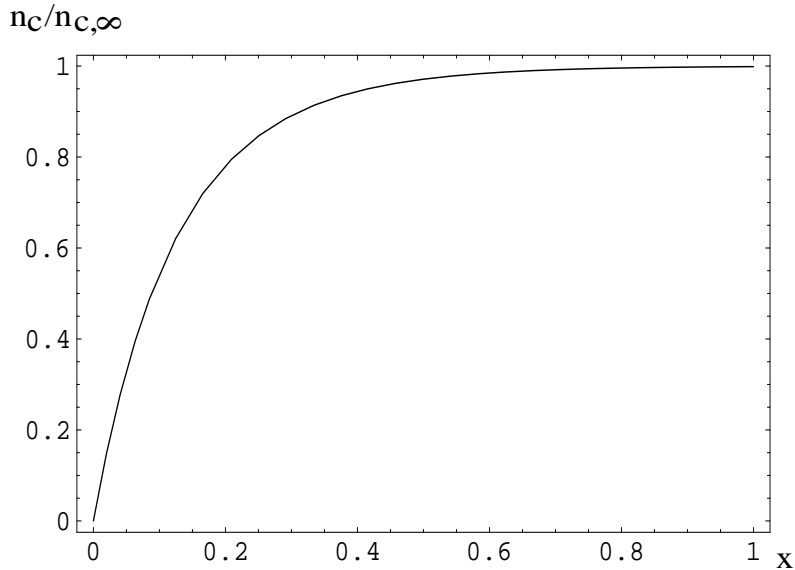}
\end{center}
\caption{Density of particles, which reside in the condensate, as a 
function of the number of e-foldings $x$ as counted from the onset of condensation.   
\label{cond}}   
\end{figure}
Saturation of the condensate density $n_c$ 
is reached within a small fraction of an e-folding at a 
value comparable to $n_0$. Hence, given an initial chemical potential 
the number of e-foldings required for a complete 
condensation of PNGB particles 
is practically determined by the critical temperature $T_0$ for 
the onset of the condensation. This shows that the assumptions 
$m_{\tiny{\mbox{PNGB}}}\sim T_0\sim \mu_e$  are consistent with the 
requirement that BE condensation be 
completed after a number of e-foldings 
which is considerably smaller than 60. From Eqs.\,(\ref{n0j}),(\ref{nc0}) 
it is seen that the asymptotic condensate density is 
\eqb
\label{acond}
n_{c,\infty}\sim f_G\frac{2.612}{e}(2\pi)^{-3/2}m_{\tiny{\mbox{PNGB}}}^3
\sim 0.11f_G\,m_{\tiny{\mbox{PNGB}}}^3\,,
\eqe
where the factor $f_G$ is determined by the gauge group $G$ and 
the result of PNGB emission. For example, if $G=SU(2)$ then 
charged PNGB, composed of pairs of $F$ and $\bar{F}$
fermions, would live in the adjoint 
representation. Assuming equal chemical 
potentials $\mu_e^a=m_{\tiny{\mbox{PNGB}}}\,,\,(a=1,2,3)$ for each charge in the gauge we 
start with would yield $f_{SU(2)}=\sqrt{3}$. 
The condensates of charged PNGB would act as an 
adjoint Higgs field $\phi^a\,,\,(a=1,2,3)$ 
and break $SU(2)$ spontaneously. We will 
address this concern in Section 3.3. To obtain the total 
number density of PNGB residing in the 
condensates the expression in Eq.\,(\ref{acond}) 
has to be multiplied by $N_F^2$.

Let us now consider case (b). Here we have 
\eqb
\label{T0}
T_0=\frac{2\pi}{m_{\tiny{\mbox{PNGB}}}}\,\exp[-2x]\,\left(\frac{1}{2.612\times 2\pi^2}
\int_{0}^\infty dp\,
p^2\left[\frac{1}{\exp[\frac{2\pi}{\La}\sqrt{p^2+m_{\tiny{\mbox{PNGB}}}^2}]-1}\right]\right)^{2/3}\,.
\eqe
We assume again that BE condensation sets in at $T_0\sim m_{\tiny{\mbox{PNGB}}}$. 
The corresponding number of e-foldings 
is $\sim 12$. As for the rapidity of the process of complete condensation 
we expect a similar behavior as for 
case (a). Setting $f_G=1$, the asymptotic condensate density is 
again $N_F^2$ times the value estimated in Eq.\,(\ref{acond}).

\subsection{Simplest field theoretic description of condensates}

Similarily to Ref.\,\cite{BarLiddle}, where the simplest curvaton model is discussed, 
we describe the flavor-sum over BEC's of PNGB in terms of one 
effective, condensed real scalar field $\sigma^*$ which is one of the 
minima of some potential $V(\sigma)$. 
Since, according to what we assumed in the last section, 
there is no scale hierarchy in the condensates, the contribution to the potential energy of 
$\sigma^*$ is $\sim (\sigma^*)^4$. Subdominant fluctuations $\delta\sigma$ give 
a contribution $\frac{1}{2}m_\sigma^2\,\delta\sigma^2$ to lowest-order, 
where $m_\sigma^2\equiv\pd^2_\sigma V|_{\sigma=\sigma^*}$. Hence we have
\eqb
\label{sigma}
V_\sigma=(\sigma^*)^4+\frac{1}{2}\,m_\sigma^2\,\delta\sigma^2\,.
\eqe
To estimate the value of $\sigma^*$ we have to 
know that the transition to the condensed phase has no latent heat, that is, there is no 
discontinuity in the energy density \cite{deGroot}. Therefore, 
the energy density $\rho$ for a single species of PNGB is given as\footnote{Because of the existence of a 
condensate of charged PNGB and consequently the existence of 
a finite $G$ field strength there is a small contribution to the energy density from 
the gauge field energy which we omit here.}
\eqb
\label{epc}
\rho=n_{c,\infty}\,m_{\tiny{\mbox{PNGB}}}\,. 
\eqe
Combining Eq.\,(\ref{epc}) and Eq.\,(\ref{acond}) yields
\eqb
\label{epcn}
\rho=0.11\,f_G\,m_{\tiny{\mbox{PNGB}}}^4\,.
\eqe
Taking into acount flavor multiplicity and the 
group structure of $G$, the factor 0.11 in Eq.\,(\ref{epcn}) converts into a number 
of order unity such that the total energy density $\rho_{BEC}$ is given as
\eqb
\label{epBEC}
\rho_{BEC}\sim m_{\tiny{\mbox{PNGB}}}^4\,.
\eqe
Combining this with Eq.\,(\ref{sigma}), we have
\eqb
\label{sigma*}
\sigma^*\sim m_{\tiny{\mbox{PNGB}}}\,. 
\eqe
This can be converted into 
\eqb
\label{sigma*,GOR}
\sigma^*\sim\sqrt{\La m_F}\,. 
\eqe
when invoking large $N_C$ counting and the Gell-Mann-Oakes-Renner (GOR) relation 
\eqb
\label{GOR}
m_{\tiny{\mbox{PNGB}}}^2\,f_{\tiny{\mbox{PNGB}}}^2\sim-m_F\,\la\bar{q}q\ra\,,
\eqe
where $\la\bar{q}q\ra$ denotes the order parameter for spontaneous 
chiral symmetry breaking. More explicitly, 
Eq.\,(\ref{sigma*,GOR}) can be
obtained as follows. On the one hand large $N_C$ counting implies that 
$f_{\tiny{\mbox{PNGB}}}\sim\sqrt{N_C}\,\Lambda$. Here $\Lambda\propto N_C^0$ 
indicates the asymptotically constant in $N_C$ scale parameter of 
the $SU(N_C)$ gauge dynamics with coupling 
constant $g_s/\sqrt{N_C}$ \cite{Manohar}. On the other hand we have 
$-\la\bar{q}q\ra\sim N_C\,\Lambda^3$. Since $m_F$ does not scale with 
$N_C$ we hence conclude that $m_{\tiny{\mbox{PNGB}}}\sim \sqrt{m_F\La}$ which 
links Eq.\,(\ref{sigma*}) and Eq.\,(\ref{sigma*,GOR}).

\subsection{Estimate of fermion mass from scale invariance of 
Gaussian isocurvature perturbations}

We now discuss the generation of Gaussian 
isocurvature perturbations due to quantum fluctuations 
of the BEC of PNGB during inflation with the 
simplifying assumption of an effective description 
in terms of a single, real curvaton field $\sigma$. The case of 
non-Gaussianity is not investigated in this paper.  

For the generation of Gaussian isocurvature 
fluctuations during inflation the 
typical perturbation $\delta\sigma=H_i/2\pi$ acrued 
in a Hubble time must be small as compared 
to the mean-field $\sigma^*$ as 
it is given in Eq.\,(\ref{sigma*,GOR}). 
Therefore, we demand
\eqb
\label{pertsigma}
\frac{4\pi^2{\sigma^*}^2}{H_i^2}\sim 4\pi^2\,
\frac{M_P^2}{\Lambda^2}\,\frac{m_F}{\Lambda}\gg 1\,,
\eqe
where $M_P$ denotes a reduced 
Planck mass defined as $H^2\equiv M_P^{-2}\,\rho$. It is natural to assume that the mass $m_\sigma$ 
of the fluctuations $\delta\sigma$  is comparable to $m_{\tiny{\mbox{PNGB}}}$. 
Hence, for $\delta\sigma$ to be the dominant cause of density perturbations and for 
the generation of a scale invariant spectrum 
of classical perturbations we 
have $m_{\tiny{\mbox{PNGB}}}^2\sim m_\sigma^2\ll H_i^2$. This is equivalent to
\eqb
\label{masssigma}
\frac{\Lambda^2}{M_P^2}\,\frac{\Lambda}{m_F}\gg 1\,.
\eqe
Hence, Gaussianity and 
scale invariance of the spectrum imply
\eqb
\label{Gs}
\frac{\La^2}{M_P^2}\gg \frac{m_F}{\La}\gg \frac{1}{4\pi^2}\,\frac{\La^2}{M_P^2}\,.
\eqe
In particular, assuming $M_P^2/\Lambda^2\sim 10^{11}$ as 
in Ref.\,\cite{HofKeil}, we 
have 
\eqb
\label{partGs}
10^{-13}\ll \frac{m_F}{\La}\ll 10^{-11}\,,
\eqe
which is a tight constraint on $m_F$. Taking 
$\frac{m_F}{\La}\sim 10^{-12}$, corresponding to $m_F\sim
10\,$GeV, yields the value $m_{\tiny{\mbox{PNGB}}}\sim 10^{-6}\,\La$ 
which is used in Section 2.2. 

During inflation the Hubble paramter is given as $H_i\sim\frac{\La}{M_P}\La$. 
From Eq.\,(\ref{Gs}) it then follows that
\eqb
\label{curvden}
\left.\frac{\rho_{BEC}}{\rho_{tot}}\right|_{\tiny{\mbox{inflation}}}\sim \left(\frac{\La}{M_P}\right)^4\,.
\eqe
This is an extremely small number. Therefore, fluctuations of $\sigma$ during 
inflation solely induce isocurvature perturbations. 

\section{Curvature perturbations and the decay of PNGB}

\subsection{The curvaton mechanism}

In this section we briefly review the 
curvaton mechanism as it was 
proposed in Ref.\,\cite{curvaton} 
and reconsidered by Lyth and 
Wands in Ref.\,\cite{LythWands} (see also \cite{Lyth}). 
We also propose a modification concerning 
the nature of the curvaton matter after inflation.  

Most economically the generation of primordial curvature perturbations 
is attributed to the quantum fluctuations of the inflaton field during inflation. 
In the standard approach fluctuations are 
adiabatically converted into classical, Gaussian curvature perturbations 
several Hubble times after horizon-exit. Upon horizon-entry after 
inflation these curvature perturbations seed the 
formation of large-scale structure.  
 
Believing in this simple picture, one ignores 
attractive models of inflation where the curvature perturbations 
cannot be generated in an adiabatic way. A generic alternative 
was discussed in Ref.\,\cite{LythWands}, and applications in more 
specific models were carried out in Refs.\,\cite{curvapps}. 
The curvaton mechanism is based on the idea that primordial curvature 
perturbations do not have to be created directly during inflation 
but can be switched on by the cosmic evolution in a postinflationary, 
radiation dominated universe. Since in the framework of gauged inflation 
we have in mind an inflation scale $\La\sim 10^{13}$\,GeV \cite{HofKeil} the 
effects of gravitational waves on the generation of curvature perturbations are neglected 
in this paper. 

Let us now turn to an application of the curvaton mechanism to a fluctuating BEC of chiral PNGB. 
A practically scale invariant spectrum of Gaussian isocurvature perturbations, 
generated by a cosmologically irrelevant light field during inflation, 
may induce primordial curvature perturbations in the course of the 
cosmic evolution following inflation. 
How isocurvature perturbations are generated on super-horizon scales during inflation 
was already discussed in Section 3.4. Here we address the postinflationary conversion 
into curvature perturbations. According to (\ref{sigma}) small 
fluctuations $\delta\sigma\ll\sigma^*$ satisfy  
\eqb
\label{eomds}
\ddot{\delta\sigma}+3H\dot{\delta\sigma}+m_{\tiny{\mbox{PNGB}}}^2\delta\sigma=0\,.
\eqe
Coherent oscillations in $\delta\sigma$ start 
if $H$ becomes of the order $m_{\tiny{\mbox{PNGB}}}$. In a radiation dominated universe 
this happens at 
\eqb
\label{massos}
\frac{a_{\tiny{mass}}}{a_{\tiny{end}}}\sim \frac{\La}{\sqrt{m_{\tiny{\mbox{PNGB}}} M_P}}\,,
\eqe
where $a_{\tiny{end}}$ denotes the scale factor at the end of inflation. With $M_P/\Lambda\sim 0.35\times 10^{6}$ and 
$m_{\tiny{\mbox{PNGB}}}\sim 10^{-6}\La$ Eq.\,(\ref{massos}) yields 
\eqb
\label{massnum}
\frac{a_{\tiny{mass}}}{a_{\tiny{end}}}\sim 1.7\,.
\eqe
Hence, $\delta\sigma$ starts to oscillate right after inflation. 
Since in Eq.\,(\ref{sigma}) the mean field part and the coherent 
oscillations $\delta\sigma$ act like a 
cosmological constant and nonrelativistic matter, respectively, we assume 
a mean equation of state for curvatonic matter which interpolates 
between pure matter ($p=0$) and vacuum ($p=-\rho$)
\eqb
\label{eoscurv}
p_{\tiny{curv}}=-\kappa\rho_{\tiny{curv}}\,,\ \ \ \ \ \ (0\le\kappa\le 1)\,.
\eqe
Admittedly, this corresponds to a rather rough 
treatment of the curvatonic matter since the ratio between the 
nonrelativistic matter part and the condensate part does evolve in time. 
The equation of state Eq.\,(\ref{eoscurv}) thus 
is merely assumed to be a time average over 
the exact equation of state taken from the 
end of inflation until curvaton decay. 
 
Continuity implies that 
\eqb
\label{cntsc}
\rho_{\tiny{curv}}=
\rho_{\tiny{curv},\tiny{end}}\left(\frac{a}{a_{\tiny{end}}}\right)^{3(\kappa-1)}\,.
\eqe
Since ``baryons'' and heavy ``mesons'' 
are of mass $N_C\,\La,\sim\La\gg m_{\tiny{\mbox{PNGB}}}$ \cite{Hof} the transition to 
the confining, $Z_{N_C}$ symmetric phase, 
which terminates inflation, should mainly generate PNGB 
radiation due to the much larger phase space. If we assume the distribution of 
this radiation to be of the Bose-Einstein type then an extra contribution $\propto \delta({\bf p})$ 
has to be added to this distribution to take 
into account the BEC of PNGB (curvaton field) already generated 
{\sl during} inflation. Consequently, the spontaneous 
breaking of chiral symmetry during 
inflation contributes to the initial conditions at 
reheating which were postulated in Ref.\,\cite{Pastor}. 

Immediately after inflation the universe is dominated by PNGB radiation. 
We will show in Section 4.3 
that this situation is maintained until the decay of the curvaton. 
Radiation as well as curvatonic matter separately satisfy 
continuity equations
\eqb
\label{conti}
\dot{\rho_\beta}=-3H(\rho_\beta+p_\beta)\, \ \ \ (\beta=\mbox{rad}, \mbox{curv})\,.
\eqe
Each fluid generates curvature perturbations $\xi_\beta$ which are 
defined on slices of uniform $\rho_\beta$ and which are time 
independent on super-horizon scales. 
From Eq.\,(\ref{conti}) we have 
\eqb
\label{xial}
\xi_\beta=-H\frac{\delta\rho_\beta}{\dot{\rho_\beta}}=
\frac{1}{3}\frac{\delta\rho_\beta}{\rho_\beta+p_\beta}\,.
\eqe
The total curvature perturbation $\xi$ is time dependent and reads \cite{Lyth}
\eqb
\label{xit}
\xi(t)=\frac{\rho_{\tiny{rad}}(t)\xi_{\tiny{rad}}+\rho_{\tiny{curv}}(t)\xi_{\tiny{curv}}}
{\rho_{\tiny{rad}}(t)+p_{\tiny{rad}}(t)+\rho_{\tiny{curv}}(t)+p_{\tiny{curv}}(t)}
=\frac{1}{3}\ \frac{r(t)}{4+r(t)(1-\kappa)}\ \frac{\delta\rho_{\tiny{curv}}}
{\rho_{\tiny{curv}}}\,,
\eqe
where $r(t)\equiv\frac{\rho_{\tiny{curv}}(t)}{\rho_{\tiny{rad}}(t)}$, 
and the density contrast $\frac{\delta\rho_{\tiny{rad}}}{\rho_{\tiny{rad}}}$ 
has been neglected. If we assume a value of $\kappa$ not far from 
unity then, according to Eq.\,(\ref{sigma}), the density contrast 
$\frac{\delta\rho_{\tiny{curv}}}{\rho_{\tiny{curv}}}$ is given as
\eqb 
\label{dencon}
\frac{\delta\rho_{\tiny{curv}}}{\rho_{\tiny{curv}}}\sim 
\frac{1}{2}\left(\frac{H_i}{2\pi m_{\tiny{\mbox{PNGB}}}}\right)^2\,.
\eqe
The consistency of this assumption will be shown in Section 4.3. 
The curvaton decays at a rate $\Gamma$ due to a decay of PNGB into $K$-fermions 
(see Section 4.2). After complete decay, 
which we assume happens instantaneously at $H=\Gamma$, $\xi$ is constant in time. 
Therefore, the prediction for the primordial spectrum of curvature 
perturbations (the Bardeen parameter) is
\eqb
\label{P1/2}
P^{1/2}_\xi\sim\frac{1}{6}\ \frac{r_{\tiny{dec}}}{4+r_{\tiny{dec}}(1-\kappa)}\ \ 
\left(\frac{H_i}{2\pi m_{\tiny{\mbox{PNGB}}}}\right)^2\,.
\eqe
In a radiation dominated universe $r_{\tiny{dec}}$ calculates as
\eqb
\label{rdec}
r_{\tiny{dec}}\sim r_i\left(\frac{a_{\tiny{end}}}{a_{\tiny{dec}}}\right)^{3\kappa+1}\sim 
\left(\frac{m_{\tiny{\mbox{PNGB}}}}{\La}\right)^4\left(\frac{a_{\tiny{end}}}
{a_{\tiny{dec}}}\right)^{3\kappa+1}\,.
\eqe
The measured value of the Bardeen 
parameter is $P^{1/2}_\xi\sim 5\times 10^{-5}$.

\subsection{Decay channels}

The ``hadronic'' modes not being PNGB, 
which exist after inflation, 
are expected to possess mass of order 
$\Lambda,N_C \Lambda\gg m_{\tiny{\mbox{PNGB}}}$ \cite{Hof}. 
Therefore, the decay of PNGB 
into these modes is kinematically 
impossible even though these ``hadrons'' may be 
charged under the weakly coupled 
$G$ sector. The only possible decays of PNGB are: (a) 
decay of charged and neutral PNGB into a pair of $K$-fermions and 
(b) anomaly mediated decay of neutral PNGB into a pair of $G$ gauge bosons.

Let us first consider case (a). We do not consider a spontaneous breakdown of 
$G$ gauge symmetry by the presence of charged BEC of PNGB. 
Using the results of the next section it will be 
argued below that this assumption can be rendered consistent. 
Therefore, the corresponding 
gauge fields are taken to be massless. 
Since we do not want to make any specific assumptions about the 
structure of $G$ we treat group factors 
to be of order unity -- in accord with 
the general level of precision intended in 
this paper. Since $K$ is assumed to be 
much smaller than $N_C$ we neglect 
flavor degeneracies. 

We assume $F$-fermion-gauge-field vertices to be parity 
conserving in the underlying $SU(N_C)$ theory. 
Therefore, the PNGB only couples to the axial current $j^5_\mu$. 
As usual, this coupling is parametrized 
by a decay constant $f_{\tiny{\mbox{PNGB}}}$ as
\eqb
\label{fpi}
\la\,0|j^5_\mu|\mbox{PNGB},p\ra=if_{\tiny{\mbox{PNGB}}} p_\mu\,.
\eqe
Diagrammatically, the decay of a PNGB into a pair of $K$-fermions is 
shown in Fig.\,2. 
For the decay of PNGB into $K$-fermions we distinguish two extreme cases.\vspace{0.2cm}\\ 
1) There is an exact flavor symmetry, 
that is, all $K$-fermions have equal mass $m_F$, 
but the $G$-gauge-boson-fermion vertex is of the A-type, $g W_\mu j_\mu^5$, and hence it 
violates parity maximally in the Lagrangian.\vspace{0.2cm}\\ 
2) The gauge-boson-fermion interaction is of the parity conserving $V$-type, 
$g W_\mu j_\mu$, but there is 
a nontrivial CKM matrix which mixes $K$-flavors of different 
mass with no large hierarchy. For a specific decay we denote 
the typical mass differences between $K$-fermion and $K$-antifermion 
by $\Delta m_F$ and the sum of masses by $\overline{m}_F$. Moreover, we assume 
the elements of the mixing matrix to be of order unity.\vspace{0.2cm}\\ 
The corresponding decay rate can be calculated on tree-level 
in analogy to the pion decay rate \cite{HalzenMartin} replacing the fermi coupling as 
$\frac{G}{\sqrt{2}}\to \frac{g^2}{m_{\tiny{\mbox{PNGB}}}^2}$ and considering a finite neutrino mass. 
According to the assumptions made under 1) and 2) instead of a $(V-A)$-type vertex we 
separately consider $A$-type and $V$-type vertices. 

For case 1) the decay rate $\Gamma_A$ can be expressed in terms of 
the gauge coupling $g$, the mass $m_{\tiny{\mbox{PNGB}}}$, 
the decay constant $f_{\tiny{\mbox{PNGB}}}$, and the mass $m_F$ of $K$-fermions as
\eqb
\label{GaA}
\Gamma_A=\frac{g^4}{2\pi}\,\frac{f_{\tiny{\mbox{PNGB}}}^2\,m_F^2}
{m_{\tiny{\mbox{PNGB}}}^3}\left(1-4\,\frac{{m_F}^2}{m_{\tiny{\mbox{PNGB}}}^2}\right)^{1/2}
\left(1+4\,\frac{m_F^2}{m_{\tiny{\mbox{PNGB}}}^2}\right)\,.
\eqe
\begin{figure}
\begin{center}
\leavevmode
\leavevmode
\vspace{7cm}
\includegraphics{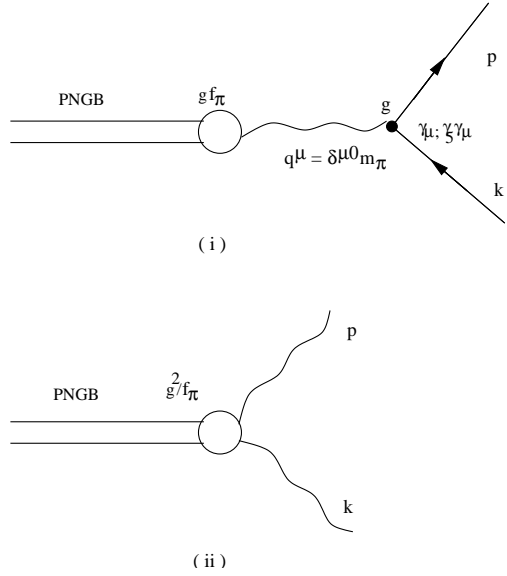}
\end{center}
\caption{Amplitudes for a charged PNGB decaying into (i) a pair of $K$-fermions and 
(ii) into two $G$-gauge bosons.
\label{PNGBdecay}}   
\end{figure}
\noindent For case 2) the expression of the decay rate $\Gamma_V$ 
in terms of the gauge coupling $g$, the mass of PNGB $m_{\tiny{\mbox{PNGB}}}$, 
the decay constant $f_{\tiny{\mbox{PNGB}}}$, and $\Delta\,m_F$, $\overline{m}_F$ reads 
\eqb
\label{GaV}
\Gamma_V=\frac{g^4}{8\pi}\,\frac{f_{\tiny{\mbox{PNGB}}}^2\,\Delta m_F^2}
{m_{\tiny{\mbox{PNGB}}}^3}\left(1-\frac{\overline{m}_F^2}{m_{\tiny{\mbox{PNGB}}}^2}\right)\left(1-\frac{\Delta m_F^2+\overline{m}_F^2}{m_{\tiny{\mbox{PNGB}}}^2}+
\frac{\Delta m_F^2\overline{m}_F^2}{m_{\tiny{\mbox{PNGB}}}^4}\right)^{1/2}\,.
\eqe
Since $\frac{m_F}{m_{\tiny{\mbox{PNGB}}}},\frac{\Delta m_F}{m_{\tiny{\mbox{PNGB}}}},\frac{\overline{m}_F}{m_{\tiny{\mbox{PNGB}}}}\ll 1$ 
we will ignore the small 
corrections in the brackets of Eqs.\,(\ref{GaA},\ref{GaV}) in the following. 
Because there is no large hierarchy in the masses 
of $K$-fermions for case 2) we will treat $\Delta m_F$ to be of order $m_F$. 

Using the GOR relation of Eq.\,(\ref{GOR}) and large $N_C$ scaling, this yields 
\eqb
\label{NCscal}
\Gamma_{V,A}\sim 
\frac{N_C}{8\pi}\,g^4\,\sqrt{m\Lambda}\sim \frac{N_C}{8\pi}\,g^4\,m_{\tiny{\mbox{PNGB}}}\equiv\Gamma\,.
\eqe
Eq.\,(\ref{NCscal}) expresses the fact that the decay width for PNGB into $K$-fermions 
scales linearly with $N_C$ and with the fourth power of a small 
coupling constant 
$g\ll 1$.

Considering case (b), we can use the result of 
\cite{PeskinSchroeder} for anomaly mediated 
decay
\eqb
\label{Gagg}
\Gamma_{\gamma\gamma}=\frac{g^4}{4\pi}\,\frac{m_{\tiny{\mbox{PNGB}}}^3}{f_{\tiny{\mbox{PNGB}}}^2}\,.
\eqe
Appealing to Eq.\,(\ref{GOR}) and using $f_{\tiny{\mbox{PNGB}}}\sim\sqrt{N_C}\,\La$, 
we derive
\eqb
\label{GaggGOR}
\Gamma_{\gamma\gamma}=N_C^{-1}\frac{g^4}{4\pi^2}\,\frac{m_F}{\Lambda}\,\sqrt{m_F\La}\,.
\eqe
This leads to
\eqb
\label{ratioGaVAGagg}
\frac{\Gamma_{\gamma\gamma}}{\Gamma}=
\frac{2}{\pi\,N_C^2}\,\frac{m_F}{\La}\,.
\eqe
Hence, decay into $K$-fermions mediated by the coupling of the axial current 
to the PNGB is extremely favoured as compared to the anomaly mediated decay 
into two $G$ gauge bosons.  

Let us come back to the issue of spontaneous $G$ symmetry breaking by charged BEC. 
This symmetry breaking induces a mass $m_W$ for the 
$G$ gauge bosons and a mass shift $\tilde{\Delta} m_F$ for 
$K$-fermions. We expect $m_W$ and $\tilde{\Delta} m_F$ to be 
of order $g\sigma^*$ and $g^2\sigma^*$, respectively. 
On the one hand, the finiteness of $m_W$ affects $\Gamma$ through a shift 
$m_{\tiny{\mbox{PNGB}}}^2\to m_{\tiny{\mbox{PNGB}}}^2+m_W^2$ 
in the $G$ gauge-boson propagator. In Section 4.3 we find 
that values of $g$ smaller than $10^{-3}$ are allowed 
suggesting that $m_W\sim g\,\sigma^*\sim g\,m_{\tiny{\mbox{PNGB}}}< 10^{-3}\,m_{\tiny{\mbox{PNGB}}}$ 
can be neglected in the calculation of $\Gamma_{V,A}$. 
On the other hand, we have $\tilde{\Delta}\,m_F\sim g^2\,\sigma^*\sim g^2\,m_{\tiny{\mbox{PNGB}}} 
\sim g^2\,\sqrt{m_F\La}\sim g^2\,10^{6}\,m_F$. Thus, 
$\tilde{\Delta}\,m_F\sim m_F$, and 
hence we may also neglect $\tilde{\Delta}\,m_F$. For a 
sudden decay at $T\sim 10^{-6}\,\La$ (see next section) 
a thermal mass $\sim gT$ \cite{Braaten} for $G$ gauge bosons can be neglected. 
However, after thermalization the thermal mass of the 
released quasi-particle excitations of $K$-fermions 
is a factor of $\sim 10^3$ larger than $m_F$.

\subsection{How large are $g$, $a_{\tiny{dec}}/a_{\tiny{end}}$, 
and $r_{\tiny{dec}}$ ?}

In this section we combine the conversion mechanism of Section 4.1 
and the above approximate determination of 
the PNGB decay rate $\Gamma$ to estimate the gauge 
coupling $g$, $a_{\tiny{dec}}/a_{\tiny{end}}$, 
and $r_{\tiny{dec}}$ as functions of $\kappa$. 
It is assumed that the 
running of $g$ with temperature $T$ 
can be neglected. 
\begin{figure}
\begin{center}
\leavevmode
\leavevmode
\vspace{7cm}
\includegraphics{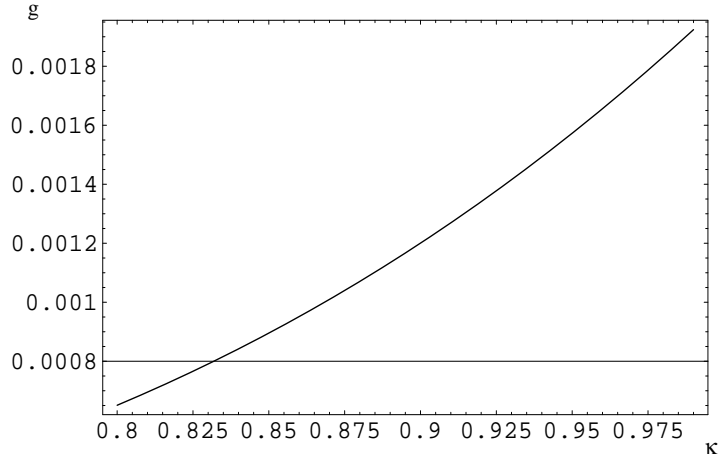}
\end{center}
\caption{The gauge coupling $g$ at PNGB decay as a function of $\kappa$. The curve 
corresponds to $P_\xi^{1/2}=5\times 10^{-5}$, $\La=10^{13}\,$GeV, 
$m_{\tiny{\mbox{PNGB}}}=10^{-6}\,\La$, and $N_C=34$. 
\label{gofkappa}}   
\end{figure}
The condition for sudden 
curvaton decay is $H=\Gamma$ \cite{LythWands}. 
Assuming radiation domination until decay and 
appealing to Eq.\,(\ref{NCscal}), 
we obtain 
\eqb
\label{adecG}
\frac{a_{\tiny{dec}}}{a_{\tiny{end}}}\sim 
\frac{1}{g^2}\ \sqrt{\frac{8\pi}{N_C}}\ \frac{\La}{\sqrt{M_P m_{\tiny{\mbox{PNGB}}}}}\,.
\eqe
Substitution of Eq.\,(\ref{adecG}) into 
Eq.\,(\ref{rdec}) yields
\eqb
\label{rdecofg}
r_{\tiny{dec}}=\frac{m_{\tiny{\mbox{PNGB}}}^{-3\kappa/2+7/2}}{g^{6\kappa+2}}\ \left(\frac{8\pi}{M_P
N_C}\right)^{3/2\kappa+1/2}\ \La^{3(\kappa-1)}\,.
\eqe
We proceed by substituting Eq.\,(\ref{rdecofg}) into 
Eq.\,(\ref{P1/2}) and by solving the result for $g$. This yields
\eab
\label{gofkaeq}
g&=&\left[\left(\frac{1}{P_\xi^{1/2}}\left(\frac{\La^2}{2\pi M_P
m_{\tiny{\mbox{PNGB}}}}\right)^2-\frac{1}{4}(1-\kappa)\right)\La^{3(\kappa-1)}
m_{\tiny{\mbox{PNGB}}}^{-3/2\kappa+7/2}\right]^{1/(6\kappa+2)}
\left(\frac{8\pi}{M_P N_C}\right)^{1/4}\,\nonumber\\ 
& & 
\eae
where $P_\xi^{1/2}$ is evaluated at curvaton decay.
In Figs.\,3, 4, and 5 graphical presentations are 
shown of Eq.\,(\ref{gofkaeq}), Eq.\,(\ref{adecG}), 
and Eq.\,(\ref{rdecofg}), respectively.
\begin{figure}
\begin{center}
\leavevmode
\leavevmode
\vspace{7cm}
\includegraphics{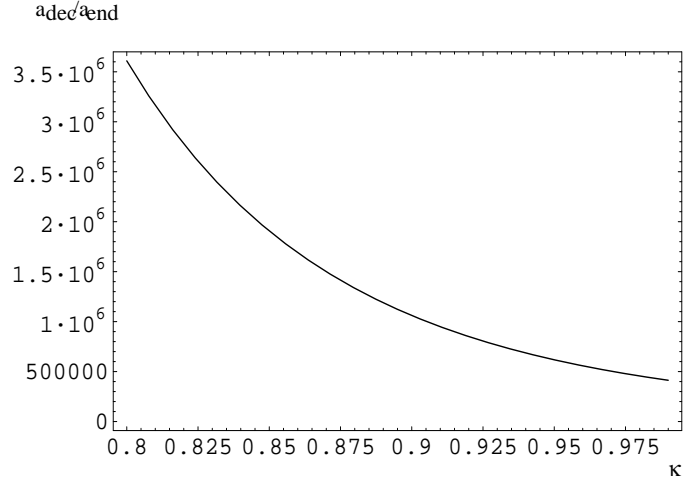}
\end{center}
\caption{The scale factor at curvaton decay 
as a function of $\kappa$. Parameter values are taken as in Fig.\,3.  
\label{aofkappa}}   
\end{figure}
\begin{figure}
\begin{center}
\leavevmode
\leavevmode
\vspace{6.5cm}
\includegraphics{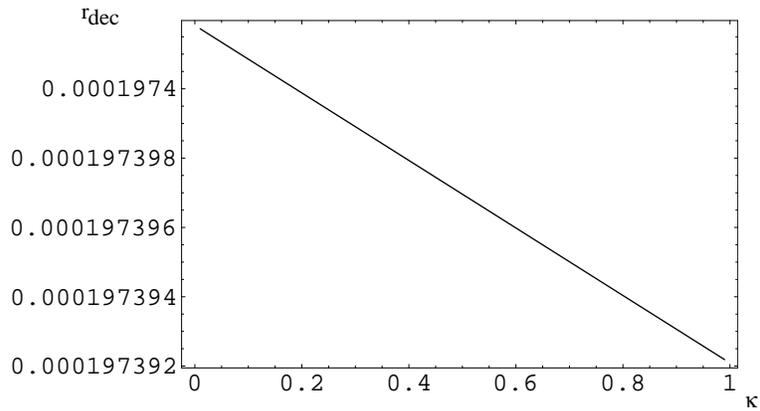}
\end{center}
\caption{The ratio of the energy densities of curvaton matter 
and PNGB radiation just before curvaton decay as 
a function of $\kappa$. Parameter values are taken as in Fig.\,3.  
\label{aofkappa}}   
\end{figure}

We are now in a position to check whether our assumptions actually are consistent 
about (i) PNGB radiation dominance at curvaton decay and 
(ii) the requirement $g<10^{-3}$ for a 
negligible mass shift of $K$-fermions due to the 
spontaneous $G$ gauge symmetry breaking by BEC. Let us first look at (i). 
On the one hand, the PNGB gas can 
be considered radiation only as long as its temperature 
$T$ does not fall below $m_{\tiny{\mbox{PNGB}}}$. At 
curvaton decay we thus demand 
\eqb
\label{radom}
\frac{a_{\tiny{dec}}}{a_{\tiny{end}}}\le \frac{\La}{m_{\tiny{\mbox{PNGB}}}}\,.
\eqe
Since $\frac{\La}{m_{\tiny{\mbox{PNGB}}}}\sim 10^6$ Fig.\,4 indicates that 
the inequality (\ref{radom}) can be satisfied for $\kappa\ge 0.9$. On the other hand, 
PNGB radiation dominates the energy density of the universe 
if $r_{\tiny{dec}}\ll 1$. According to Fig.\,5 this is well 
satisfied for the entire indicated range of $\kappa$ values. Therefore, PNGB
radiation dominance at curvaton decay is possible.  

From Fig.\,3 we see that (ii) 
is met for $\kappa\le 0.9$. Hence, 
(i) and (ii) together fix $\kappa$ to be $\kappa\sim 0.9$. 

Fig.\,5 indicates that the ratio of curvaton to radiation energy 
increased from $r_{\tiny{end}}\sim 10^{-24}$ at the end of inflation to 
$r_{\tiny{dec}}\sim 10^{-4}$ at curvaton decay. This is the feature that 
allows for a conversion of the isocurvature perturbations, imprinted in 
the curvaton field at inflation, into 
curvature perturbations after a postinflationary expansion of $\sim 10^6$.  
 
Due to the extremely small coupling $g\sim 10^{-3}$ cosmology would be
dominated by a plasma of $K$ fermions and $G$ gauge bosons after curvaton decay. 
The $K$ fermions of mass $\sim$10\,GeV are wimpy and 
therefore potential contributers to 
the cold dark matter of today's universe.

\section{Summary}

The purpose of this paper was to propose and 
test a generation and decay mechanism for a 
light field that can play the role of a curvaton 
in the framework of gauged inflation. The idea is 
that the spontaneous breaking of a 
chiral symmetry {\sl during} inflation and the 
subsequent Bose-Einstein condensation of 
the associated pseudo Nambu-Goldstone 
bosons dynamically generate such an 
effective light scalar field. 
Postinflationary decay of 
this field and its excitations into fermions is mediated by 
weakly coupled gauge dynamics. In this 
scenario the mass scales are fixed by 
chiral symmetry and the requirement 
of the Gaussianity of the isocurvature 
perturbations that are generated from 
the quantum fluctuations of the curvaton during inflation.  
  
\section*{Acknowledgements}    

The author would like to thank M. Keil, S. Pastor, and D. Semikoz 
for useful discussions. 

\bibliographystyle{prsty}

\end{document}